# A Holistic Framework for Open Low-Power Internet of Things Technology Ecosystems

Peng Hu[1], *Member, IEEE*

**Abstract**

The low-power Internet of Things (IoT) has been thriving because of the recent technological advancement and ecosystems meeting the vertical application requirements and market needs. An open IoT technology ecosystem of the low-power IoT has become increasingly important to all the players and stakeholders and to the research community. However, there are several mainstream low-power IoT ecosystems available out of industry consortia or research projects and there are different models implied in them. We need to identify the working framework behind the scene and find out the principle of driving the future trends in the industry and research community. With a close look at these IoT technology ecosystems, four major business models are identified that can lead to the proposed ecosystem framework. The framework considers the technical building blocks, market needs, and business vertical segments, where these parts are making the IoT evolve as a whole for the years to come.

## I. Introduction

Internet advances with the openness in mind, so does the Internet of Things (IoT). An IoT system benefit from various kinds of technologies and developmental efforts driven by open IoT technology ecosystems involving open standards, open source tools, and open platforms with key stakeholders. Over the past decade, the innovative sensors, embedded systems, cloud computing, wireless networking technologies have been enriching the openness of IoT systems and fulfilling the needs of IoT system development. As a result, on the one hand, these technologies enable the extremely less power consumption on IoT devices than before, which enables a broad spectrum of zero-battery and battery-powered applications. On the other hand, an IoT system can be effortlessly built with an IoT ecosystem including the off-the-shelf IoT building blocks, ranging from sensors, interface circuits, embedded modules, and connectivity modules, to cloud service and data analytical tools.

Low-power wireless networking technologies are indispensable for the low-power Internet of Things (IoT) systems ranging from wearable devices, smart appliances, smart meters, to smart city systems. There are mainstream low-power wireless networking technologies, from zero-battery and low-power wireless personal area network (LPWPAN) open standards, such as EnOcean, Bluetooth LE (BLE), 6LoWPAN, and ZigBee, to low-power wide area network (LPWAN) open standards, such as LTE-M and LoRA. These open-standard-compliant technologies are mostly driven by industry players and create the IoT technology ecosystems. Existing IoT ecosystems result in the fragmented IoT market and the effort of creating a consortium or alliance is the viable way to ensure the monetized return of the IoT businesses. This trend has divided IoT into vertical applications and divided market into vertical business segments. As a result, technology and media competition of the IoT market is no longer between individual firms but "among ecosystems of firms operating in loose alliance" [1]. We need to understand these ecosystems of the low-power IoT business.

---

[1] Dr. Peng Hu was with CMC Microsystems, Kingston, Canada and is now with School of Information & Communications Technology, Seneca College of Applied Arts and Technology, 70 The Pond Rd., Toronto, Ontario, Canada M3J 3M6
Email: penghu@ieee.org

However, the current research in the literature mainly discusses the business ecosystem of IoT arising from the supply network [2]–[4] without the consideration of an IoT technology ecosystem involving all IoT players and technical elements with an updated view of technology ecosystems, addressing the IoT research and industry trends, and identifying the framework behind the scene. In this way, we need to have a holistic framework identifying what the key technical building blocks are, how they interacts with business segments, and how to make the IoT business, enablers, stakeholders, and market integrate and co-evolve, in order to generate promising social and economic impact in the future.

In this article we provide an updated overview of the IoT ecosystems, discuss the models of open low-power IoT ecosystems, and propose a holistic framework that allows IoT systems from business and technology perspective to evolve and to integrate.

In the following sections we present the concepts of IoT ecosystem, an overview of the existing IoT ecosystems categorized by four business models, as well as a framework that can provide a holistic view of the major low-power IoT ecosystems. Then, we discuss the case studies of the two successful ecosystems, AllJoyn and ARM mbed, with a conclusion.

## II. Related Work

There are some related works discussing about the related topics about IoT ecosystems. We firstly introduce the IoT ecosystems and related business models. Then, the classical IoT-based business ecosystems are reviewed in contrast to the low-power IoT technology ecosystems, followed by the common features of IoT ecosystems.

In [5], IoT ecosystems can be represented by the three components: enabling technology, IoT viable marketplace, and applications desirable to users or stakeholders. The IoT Architectural Reference Model (ARM) [6] has started an effort on creating an architectural model that allows IoT devices to interoperate in a standard way supported by the use cases introduced in [7]. The players in the IoT market is mentioned in [4] including commercial players, research and academia, governments and utilities, and others.

The business ecosystem was introduced in [2] where the key players in addition to the ones in a supply chain are considered such as universities, industry associations and other stakeholders. An IoT-based business ecosystem is introduced in [3], where the and the technical RFID ecosystem is mentioned in [8]. We should note that the IoT ecosystem in this article is not a business ecosystem, but a technology ecosystem in consideration of all elements and business models.

A business model is an element in any IoT ecosystems because the ecosystems are created to generate an economic return in the end. The term of business model has different definitions [9] according to contexts. The new IoT-based value creation in the IoT business model is mentioned in [10], where the interoperability of products and services as smart thermostats and light bulbs can generate new services along the data flow, such as processing optimization and forecasting. The business model framework that identifies the components of developing a business model for companies is discussed in [11], where the value proposition is considered the most important element of a business model. In addition to value creation, adoption of IoT can also create the operational process improvements, cost reduction, and risk minimization [3].

There are common features of IoT technology ecosystems. One feature is they need to have technical IoT building blocks such as the hardware platforms, operating systems (OSes), and software frameworks. Technical IoT building blocks play an important role in the IoT technology ecosystem and they mostly meet the open standards. There are rich sets of the technical IoT building blocks from the chip level, infrastructural level, to the end-user application level. These open standards may include open specifications by international consortia or standards made by international standardization bodies. Open standards make sure the interconnectivity and interoperability across IoT systems and de-risk the development for new products and services. This is welcomed by most of the business entities. For example, the IoT-A ARM architecture [6] necessitate different standards for different hardware/software components of a system which they need to be compliant with at different layers following the OSI model. Because of the versatility of IoT applications, various standardization efforts have been made by IEEE, IETF, ITU, and IEC, as well as industry consortia, including AllJoyn Alliance, Bluetooth Special Interest Group (SIG), ZigBee Alliance, Z-Wave Alliance, Open Connectivity Foundation (OCF, formerly called OIC), and EnOcean Alliance. Most standardization efforts made by industry consortia are on top of the underlying component standards and they provide the software frameworks of fast-prototyping an IoT system with inherent interconnectivity and interoperability of other systems with the same software framework.

### III. Overview of the Current Low-Power IoT Technology Ecosystems

An IoT ecosystem can be further split into sub ecosystems and a way of identifying these sub ecosystems are through the technical building blocks, although not all the building blocks have their own ecosystem model.

Table 1 shows the essential parameters in comparison with different low-power IoT ecosystems from different industry consortia or projects. We name ecosystem after the industry consortium or project name. Where the IoT ecosystem with open standards, proprietary standards such as ANT are not included. The license is listed for each ecosystem to show the openness and the supported OSes are listed as well. The vertical application of the ecosystem shows the business foci of each ecosystem. Most of the ecosystem identifies the home automation, including appliances, lighting, climate control, energy management, access control, safety and security in a home.

The energy harvesting based EnOcean IoT ecosystem is shown in Table 1, where it targets at the home automation applications. EnOcean ecosystem is built on top of ISO/IEC 14543-3-10 standard, where it fits a new market, connects the electronics/semiconductor industry and provides development tools.

The LPWPAN includes Bluetooth (and BLE defined in the 4.0 specification), ZigBee, DASH7, and Z-Wave. They originally offered specifications based on the physical-and link-layer open standards such as IEEE 802.15.4, ISO/IEC 18000-7, and IEEE 802.15.1, and now provide a full-fledged development kits and tries to expand the established markets in home automation, to the markets of retail or industrial applications. We can see that they all have similar licensing model where use of software libraries are free but the final products need to go through the licensing or certification process. In addition, WirelessHART is a special ecosystem built on top of IEEE 802.15.4-compliant radios, where it targets the manufacturing automation vertical application that differs itself from other LPWPAN technologies.

The rising of low-power wide area network (LPWAN) technologies can extend the communicate coverage to the city wide which fill in the market gap left by the LPWAN technologies. LET-M and LoRa are two examples based on open standards/specifications. LTE-M is alongside the LTE infrastructure so that the LTE carrier providers would welcome this, while LoRa supports the private network without the LTE infrastructure. The licensing model of LTE-M is similar to LTE and it is expected to be bound to the platform providers, while LoRa adopts the LGPL software license.

The ecosystems such as IoTivity, Thread, AllJoyn, and ARM mbed can work with the underlying low-power transport technologies such as BLE, 6LoWPAN, and IEEE 802.15.4. IoTivity and AllJoyn are Linux Foundation software framework projects where they have differences. They have different architectures and AllJoyn focuses on the service framework, device libraries, and applications, while IoTivity focuses on the device discovery, transmission and management. ARM mbed supports low-power radios and focus on the ARM-based hardware platforms with the support from firmware, hardware design, middleware, and cloud services.

In addition, it is important to realize the vertical segments that an IoT software framework can address. Although most of them are in the home automation, smart home, and smart city business. However, there is possibility of the IoT software frameworks to be used in other segments such as manufacturing systems, energy systems, and automotive systems.

| IoT Ecosystem | Model | Software License | Supported OSes | Connectivity Technology Supported | Security Support | Firm / Alliance | Vertical Applications |
|---|---|---|---|---|---|---|---|
| Thread | IV | 3-Clause BSD-based license (OpenThread by Nest) | Platform independent | IEEE 802.15.4/6LoWPAN | Yes | Thread Group | Home automation |
| IoTivity | IV | Apache License 2.0 | Android, Tizen, Arduino, native Linux, and platform-independent library | BLE/Bluetooth, WiFi Direct, Ethernet | Yes | Open Connectivity Foundation (formerly OIC) | Generic |
| AllJoyn | IV | Internet Software Consortium (ISC) | Windows, Android, Linux, iOS, and platform-independent library | WiFi, Serial, Power Line Communication (PLC), Ethernet, 6LoWPAN, EnOcean | Yes | AllJoyn Alliance | Home automation |
| ARM mbed | I, II | Apache License 2.0 (and other licenses) | Platform-independent library | Ethernet, WiFi, 6LoWPAN / Thread, BLE | Yes | ARM | Generic |
| ZigBee | I, III | Certification process is required | Platform-independent | IEEE 802.15.4/ZigBee (can be integrated to AllJoyn and OCF) | Yes | ZigBee Alliance | Smart home, lighting, utility, and retail industry |
| Z-Wave | I, III | Certification process is required | Platform-independent | ITU-T G.9959/Z-Wave (can be integrated with AllJoyn and OCF, and HomeKit) | Yes | Z-Wave Alliance | Wireless home control and monitoring |
| EnOcean | I, III | Certification process is required | Platform-independent | ISO/IEC 14543-3-10:2012 | Yes | EnOcean Alliance | Home automation |
| WirelessHART | III | Certification process is required | Platform-independent | IEC 62591 | Yes | HART Foundation | Manufacturing automation |
| LoRa | I, III | LGPLv2.1 | Platform-independent | LoRaWAN specification | Yes | LoRa Alliance | Generic |
| LTE-M | I | N/A | Platform-independent | LTE | Yes | 3GPP | Automotive, smart grid, smart city, healthcare, smart home, industrial |
| DASH7 | I, III | OpenTag License | OpenTag OS | ISO/IEC 18000-7/DASH7 Mode 2 | Yes | DASH7 Alliance | Building automation, smart energy, advertising, automotive, logistics |
| Bluetooth | I | Licensing process is required | Platform-independent | IEEE 802.15.1/Bluetooth/BLE | Yes | Bluetooth SIG | Consumer electronics, healthcare |
| RIOT | I, IV | LGPLv2.1 | Linux, platform-independent | IEEE 802.15.4/6LoWPAN, WiFi | Yes | N/A | Generic |
| Contiki OS | I, IV | 3-Clause BSD-based license | Contiki OS | IEEE 802.15.4/6LoWPAN | Yes | N/A | Generic |

**Table 1. List of major low-power IoT technology ecosystems where the key parameters are listed including the type of business model, software license, supported operating systems (OSes), security, alliance, and vertical applications.**

## IV. Business Models of Low-Power IoT Technology Ecosystems

Here we inherit the definition in [12] to define the business model of IoT ecosystems here as the model that the value of the ecosystem consisting of a consortium or an alliance offers to all the key players including member firms, customers, developers, and researchers. There are four typical business models behind the aforementioned ecosystems as shown in Table 1.

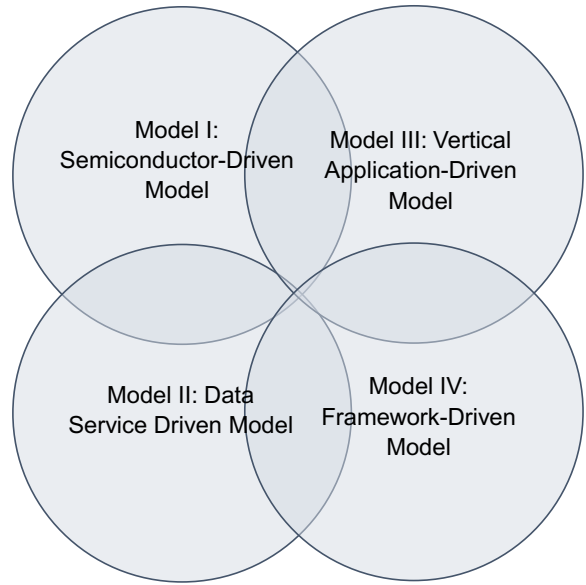

Figure 1. Four common models of low-power IoT ecosystems

### A) Model I: Semiconductor-Driven Model

This model is created by the hardware-based ecosystems from semiconductor business models. The semiconductor companies manufacture the hardware components and creates the software libraries and development kits with the partners for customers including developers and system integrators. The research element is the important enabler for design and fabrication process innovations reflected by the Moore's Law. This model is alongside the traditional vertical segment of embedded systems where the main customers of semiconductor companies are not end users but companies who will buy the hardware components in a large volume to use them in the final products. In recent years, the huge opportunity and availability of IoT technologies have further resulted in the additional competition among the horizontal semiconductor segments, where they are now competing to provide more open, easy-to-use and high-quality modules and development kits than before. For example, in 2014, Freescale released the support for MEMS Industry Group by offering loyalty-free open-source sensor fusion library that allows the easy development IoT application based on the essential sensor fusion building blocks. In 2015, Texas Instruments (TI) released the multi-protocol SoC CC2650 chip with a complete software/hardware development kits, such as software libraries, mobile app examples, and cloud service connectors, in addition to the standard offerings such as hardware reference design and application notes. ARM mbed follows a typical semiconductor-driven model where the various hardware platforms and tool chains by ARM mbed are based on ARM microprocessors. Although this model is well validated by all major semiconductor companies such as TI, Atmel, ST, NXP, and Freescale, but ARM mbed has built a full-featured technical ecosystem connecting hardware vendors and providing the technical building blocks from chip-level to open circuits, OS, middleware, and cloud services.

## B) Model II: Data Service-Driven Model

The data-driven model aims to provide the security, messaging, schemas, processing, and interoperation of generic data required by the IoT applications. It is undeniable to say data is the main part of the value creation in many businesses, and the ecosystem built with this model has been validated by the market in alignment with the market success of the existing platforms such as Amazon IoT platform, IBM Bluemix, Google Cloud Platform, IFTTT, and ThingSpeak. This data service can be provided independently as long as the open data connectors such as REST data connector are available on the hardware platform. In addition, the developers and researchers as the possible customers have already advanced and been benefited from the data service platforms.

## C) Model III: Vertical Application-Driven Model

Vertical applications are to denote particular application areas. This vertical application-driven model refers to the IoT application sets that differ from each other, where each would require unique specifications. For example, in addition to the LoRa, ZigBee, Z-Wave, DASH7, and WirelessHART shown in Tabl1, the LORD MicroStrain platform provides another example ecosystem which focuses on the industrial sensing applications. On the sensor device it has specialized protocol for microsecond time synchronization based on IEEE 802.15.4-compliant radios, and one the cloud end it provides the on-line data processing APIs. Another example is the NI wireless monitoring platform where it provides a set of development kits that can work with LabView and data acquisition chases for the industry-grade sensing applications. Shimmer provides a similar IoT solution to the healthcare applications.

An IoT system is expected to be a multi-purpose computer system that can support various types of tasks, which generates new business opportunities and values. However, when talking about the IoT ecosystem, we need to conduct an abstraction from this multifaceted use cases of IoT. From a system's perspective, an IoT system can be split into few main sub systems from sensor and actuator systems, to embedded systems, and distributed systems. From a computer network's perspective, it can be split into the classical OSI layers from physical layers, network and transport layers, to application layers.

## D) Model IV: Framework-Driven Model

The framework-driven model refers to the offering of software frameworks working at a high level of an IoT system with the middleware that makes it hardware platform agnostic. The example of this business model include AllJoyn, Thread, and IoTivity, which are software framework based ecosystems.

Today's IoT system may employ one or more transport technologies and it is important a software framework should support them. For example, a wireless IoT router needs at least a wireless transport and Ethernet connectivity in order to communicate within the local wireless network and to the external IP backbone network. In order to support flexible number of transport technologies, we need to make sure they can be "bridged" to these networks. With the software frameworks, these networks can be integrated into the application protocols in the framework.

Figure 1 shows the possible overlaps between aforementioned models which is true from Table 1 that an ecosystem can fit more than models. For example, it is possible that a chip manufacture provides a full set of software development kits that offer some features in the framework-driven model.

Moreover, one may wonder the model implied in the IoT ecosystems such as Arduino, Raspberry Pi, or Beagle Bone. They indeed have enabled the proliferation of IoT applications for years, but since they are not focused on the low-power IoT ecosystems, they are out of the scope of this article.

In addition to the aforementioned models, there exists a research-driven model that the IoT ecosystem originates from the research community. Contiki OS and RIOT are two examples of it where both of them focus on the low-power IoT solutions. There is a correlation of this model with Model I or IV as some research-driven IoT ecosystems have a well-designed business model with successful market share, support, and platform support from industry. However, the research-driven model of IoT ecosystem needs to plan for the commercialization stage in order to avoid the possible support and maintenance issues that cannot be entirely resolved by the open-source community.

## V. A Holistic Framework for Low-Power IoT ecosystems

Based on the aforementioned business models, we introduce a framework for low-power IoT ecosystems as shown in Fig. 2.

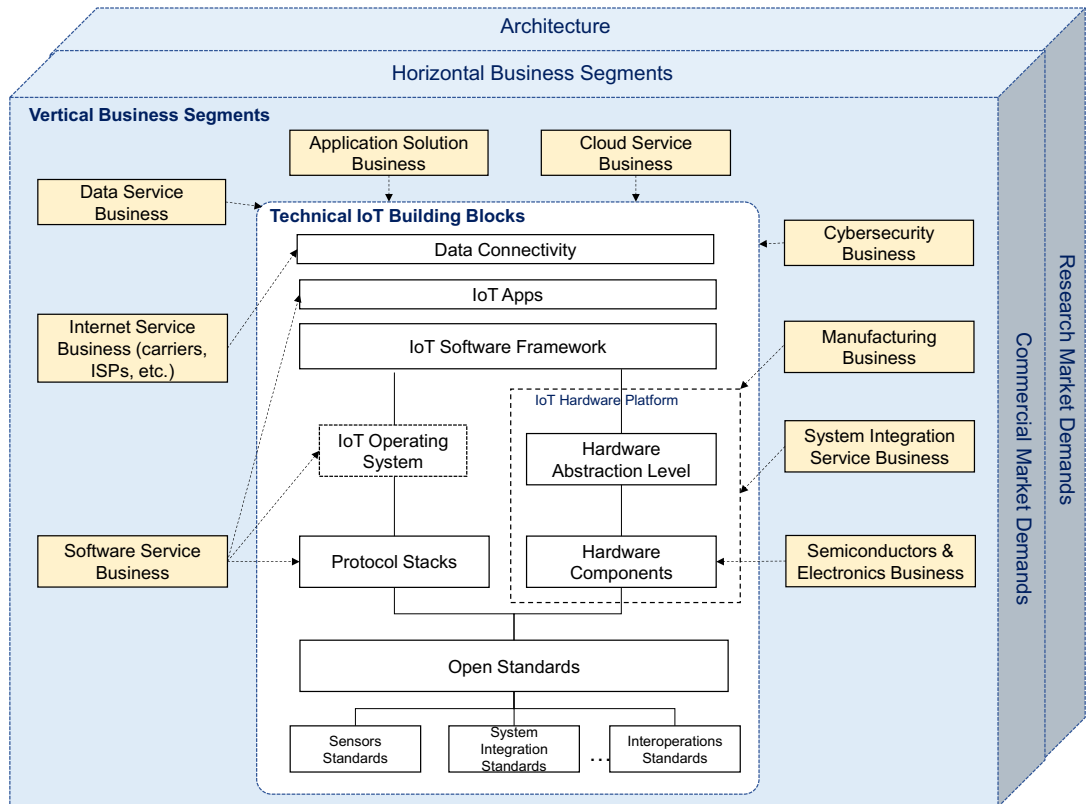

**Figure 2. A holistic framework for low-power IoT ecosystems. This framework shows the research and market demands as the driving force as well as the vertical business segments and technical IoT building blocks.**

Figure 2 contains two parts in the front plane: the technical IoT building blocks part and the common vertical business segments part. The horizontal business segments are briefly shown on the top plane of Fig. 2, where there can be multiple similar business entities of a business block shown on the front plane. The architecture plane indicates the various systems following an architecture based on the common IoT building blocks. On the side plane, the commercial and research market demands are shown as the market driving force of the elements shown on the front plane.

In the technical IoT building blocks, the open standard block connects to all other blocks including the sensor/device-level blocks such as drivers, protocol stacks and hardware platforms as well as application-level blocks such as the software framework and IoT apps where the software framework can make all the IoT building blocks work together. There are three categories of open standards, such as sensors standards, system integration standards, and interoperation standards, where each category contains a series of standards. The data connectivity block requires physical connection to provide actual wireless connection between IoT devices. These technical IoT building blocks connect the business and research worlds and are mostly open in terms of technology, licensing model, and integration with other building blocks.

The blocks representing vertical business segments sit outside the technical building blocks and the dotted arrow showing the relationship between business blocks and technical building blocks.

Semiconductor and electronics business provides the essential bedrocks of IoT systems such as sensors, actuators, and electronic components. The system integration business can provide the hardware platforms or out-of-the-box modules based on the electronic components. Manufacturing business includes any manufacturing business regarding the IoT hardware components, such as PCB fabrication and 3D printing business for the custom cases for an IoT device.

The data connectivity building block requires the Internet connection services in the "Internet service business" block. Although data connection between local devices in a LPWPAN is free, the data communication between a router and a remote service is not. A company providing LTE-M based LPWAN data connectivity service for smart city applications fits the on-premises or subscription service business model.

Data service business includes data storage, data processing, and data operations. These common services are important for many IoT applications. For example, for the smart thermostat application, all thermostat sensing data can be hosted in the data service provider's premises where the data analysis tool can process the data and show the results to the users instead of showing a large amount of data.

Software service business is related to the software related components such as protocol stacks, operating systems, software frameworks, and applications. This software business includes consulting services, outsourcing services, software support services, etc. Each of these services already has a mature business model.

The cloud service business closely relates to the modern IoT applications where the distributed services or a remote server can be accessed and hosted. The successful business models based on the service models such as Software-as-a-Service (SaaS), Platform-as-a-Service (PaaS), and Infrastructure-as-a-Service (IaaS) can be used in this business segment.

The application solution business relates to the complete IoT system where it can provide service to different kinds of IoT applications such as home automation, smart city, smart grid, and manufacturing automation. These application solutions need to be validated and well-designed and it generate unique value to the customers.

The cybersecurity business provides the security solution to the IoT system which is distributed in essence. Although we have seen in Table 1 that security features are provided with each IoT technology but it is just a basic building block which is far less possible to tackle all the cyber threats. The existing cybersecurity industry is the example of this business.

## VI. Discussion

AllJoyn and ARM mbed are the recent successful IoT ecosystems that fit the previously proposed Model I and IV and the framework. Both of them are well connected to the business stakeholders, researchers and developers. The AllJoyn software development kits (SDKs) are open to the public and its services are based on the low-power wireless open standards and provides the agent middleware running on the, computers, routers, and mobile/embedded devices. It fits the framework in Fig. 1 in that the commercial market needs a full-fledged and easy-to-use solutions where they are easily built with any possible OSes, including Android, iOS, Windows, Linux, or any generic embedded OS. AllJoyn abstracts the common business logics of the mobile based IoT applications and provides separate software libraries to the mobile phones and embedded devices while keeping the same communication protocol. The application-layer communication protocol provides an easy way to enable interoperations between any AllJoyn devices, although it may introduce the latency issue of data transmission for the time-critical application. However, this is not a concern in most of consumer-grade applications. The AllSeen Alliance requires the membership to create a new project or work on the existing projects but it provides the free members for academic users to do so.

ARM mbed provides SDKs from the firmware, OS, connector middleware, to the cloud service which can be mapped to the technical building blocks in Fig. 2. It supports a broad range of low-power IoT protocols and application-layer protocols, and it has built partnerships with hardware manufactures where many ARM-based hardware platforms can support ARM mbed. Any users can effortlessly build solutions on top of it with a low cost, although there are multiple licenses involved in the software components. The loosely coupled architecture based on open standards allow the vertical business segments to utilize the ARM mbed solutions. For example, application solution business can build a dedicated virtual application on top of the ARM mbed ecosystem.

In addition, the value creation out of AllJoyn and ARM mbed are embodied in the support for device, edge, and cloud computing. The technical building blocks provided by ARM mbed and AllJoyn enable the edge and cloud computing as well as the intrinsic on-node computing. To this end, the intelligence and data processing can be implemented within any architecture.

## VII. Conclusion

With the fast-advancing low-power wireless technologies at different scales including LPWPAN and LPWAN, the technical barrier of IoT applications is further removed, resulting in the proliferation of IoT applications and business opportunities in the coming years. The challenge now is how to identify a business model in an IoT technology ecosystem to integrate these technologies. To this end, it is important for us to review the IoT system in consideration of technology evolution, markets, and related businesses. The proposed business models and framework that reveal the principle of various low-power IoT technology ecosystems and help us understand the synergy and convergence of the key elements in the ecosystems.